\documentclass[onecolumn,aps,prd,preprintnumbers,showpacs,superscriptaddress,nofootinbib,amsmath,amssymb,floats,floatfix,showkeys,notitlepage,longbibliography]{revtex4-1}

\usepackage{orcidlink}
\usepackage{comment}
\usepackage{lipsum}
\usepackage{graphicx}
\usepackage{subfigure}
\usepackage{palatino}
\usepackage{sans}
\usepackage{hyperref}
\hypersetup{colorlinks=true,linkcolor=blue,urlcolor=blue,citecolor=blue}
\usepackage[toc,page]{appendix}
\usepackage[normalem]{ulem}
\usepackage{adjustbox}
\usepackage{latexsym}
\usepackage{amsmath}
\usepackage{amssymb}
\usepackage{amsfonts}
\usepackage{dcolumn}
\usepackage{bm}
\usepackage{tikz}
\usepackage{bigints}
\usepackage{array,tabularx,multirow,booktabs}
\usepackage[tracking=true]{microtype}
\usepackage{soul} %for highlighting
\SetTracking{}{500}
\SetTracking{encoding={*}, shape=sc}{40}
\UseRawInputEncoding %for inputenc error%
\allowdisplaybreaks

\renewcommand{\Re}{\ensuremath{\mathrm{Re}}}
\renewcommand{\Im}{\ensuremath{\mathrm{Im}}}
\newcommand{\be}{\begin{equation}}
\newcommand{\ee}{\end{equation}}
\newcommand{\bea}{\begin{eqnarray}}
\newcommand{\eea}{\end{eqnarray}}

\begin{document} \sloppy

\title{Probing Quantum Gravity in Stellar Spacetimes: Phenomenological Insights}

\author{Reggie C. Pantig
\orcidlink{0000-0002-3101-8591}
}
\email{rcpantig@mapua.edu.ph}
\affiliation{Physics Department, Map\'ua University, 658 Muralla St., Intramuros, Manila 1002, Philippines}

\author{Ali \"Ovg\"un
\orcidlink{0000-0002-9889-342X}
}
\email{ali.ovgun@emu.edu.tr}
\affiliation{Physics Department, Eastern Mediterranean
University, 99628, Famagusta, North Cyprus
via Mersin 10, Turkiye.}

\author{Gaetano Lambiase \orcidlink{0000-0001-7574-2330}}
\email{lambiase@sa.infn.it}
\affiliation{Dipartimento di Fisica ``E.R Caianiello arcsecond, Università degli Studi di Salerno, Via Giovanni Paolo II, 132 - 84084 Fisciano (SA), Italy.}
\affiliation{Istituto Nazionale di Fisica Nucleare - Gruppo Collegato di Salerno - Sezione di Napoli, Via Giovanni Paolo II, 132 - 84084 Fisciano (SA), Italy.}

\begin{abstract}
We explore the weak-field phenomenology of a compact star spacetime modified by quantum gravitational corrections derived from the effective field theoretical (EFT) approach by Calmet et al. [1]. These corrections, encoded in non-local curvature-squared terms, distinguish matter-supported geometries from vacuum solutions by contributing nontrivial modifications at order $O(G^2)$. Using the corrected metric, we analytically derive expressions for the deflection of light and time-like particles via the Gauss-Bonnet theorem. We further compute the perihelion advance of Mercury, Shapiro time delay, and gravitational redshift within this framework. Each classical observable acquires quantum corrections that, though exceedingly small—on the order of $ 10^{-9}$ arcsecond per century for perihelion precession and $ 10^{-18}$ arcsecond for light deflection—represent potential imprints of quantum gravity. The Shapiro delay and redshift likewise exhibit finite, source-dependent deviations from their general relativistic predictions due to the modified temporal metric component. While current observational capabilities remain insufficient to detect these minute effects, the analysis demonstrates that quantum gravitational signatures are embedded even in weak-field observables. Last, we study massless scalar perturbations in static, spherically symmetric spacetimes by analyzing their quasinormal modes (QNMs) and greybody factors using the WKB method and Pade resummation. Our findings demonstrate that increasing the coupling parameter enhances spacetime stability and significantly influences emission spectra through frequency-dependent transparency. Moreover, the results underscore that quantum-corrected star metrics yield phenomenological distinctions from classical black holes, particularly near the Planck scale, where vacuum solutions lose validity.

\end{abstract}

\pacs{95.30.Sf, 04.70.-s, 97.60.Lf, 04.50.+h}
\keywords{General relativity; Weak deflection angle; Shapiro time delay; Gravitational redshift; Quasinormal modes}
%only 5 keywords

\maketitle

%\tableofcontents

\section{Introduction} \label{intr}
In the realm of astrophysics, the study of stars has long been a cornerstone of our understanding of the cosmos. Recent advancements in theoretical and experimental physics, particularly in the domain of quantum gravity, have sparked intrigue regarding the potential existence of subtle corrections to the behavior of stars at the quantum level. These corrections, if present, could offer profound insights into the unification of quantum mechanics and general relativity (GR), the two pillars of modern physics. The literature concerning the study of quantum corrections applied to compact objects, such as stars, white dwarfs, and neutron stars, is vast. To cite a few, Ref. \cite{Wang:2019gry} investigates long-distance, low-energy, leading quantum corrections to the gravitational potential for scalarized neutron star binary systems. It treats GR as an EFT and considers graviton and scalar particle exchanges. The study sheds light on the quantum effects of the gravitational interactions of these compact objects. In another work \cite{Huang:2022zia}, a general framework is presented for using quantum error correction codes to protect and image starlight received at distant telescope sites. The scheme involves capturing the quantum state of light into a non-radiative atomic state via Stimulated Raman Adiabatic Passage and imprinting it into a quantum error correction code. Ref. \cite{BenAchour:2020mgu} presents an effective quantum extension of the seminal Oppenheimer-Snyder (OS) collapse, modeling singularity resolution using the effective dynamics of spatially closed-loop quantum cosmology. The bouncing star discussed here admits an IR cutoff and corresponds to a pulsating compact object based on Loop Quantum Cosmology techniques. While it cannot describe a black-to-white hole bounce, it provides a concrete realization of a pulsating compact object within the framework of loop quantum cosmology.

The quest to unify quantum mechanics with GR continues to challenge the foundations of theoretical physics. While a complete theory of quantum gravity remains elusive, significant progress has been achieved through EFT techniques, which allow one to calculate quantum corrections to classical gravitational fields at energy scales far below the Planck scale \cite{Calmet:2019eof,Calmet:2020vuh,Battista:2023iyu}. Within this low-energy regime, the gravitational field is treated as a classical background with quantum fluctuations incorporated as higher-order curvature terms. The resulting non-local effective actions preserve general covariance and offer a model-independent framework to extract observable signatures of quantum gravity.

Recent investigations have increasingly focused on the physical implications of quantum corrections to spacetime geometry in astrophysical contexts. Compact stellar objects such as neutron stars, which possess non-vanishing stress-energy, offer a natural setting where loop-induced corrections to the classical metric can arise. These corrections modify the semiclassical gravitational dynamics in ways not captured by GR alone. In contrast, vacuum solutions like eternal Schwarzschild black holes exhibit a striking insensitivity to such modifications: quantum corrections at second order in curvature vanish in these cases within EFT frameworks. This disparity reflects a fundamental distinction introduced by quantum gravity — while GR predicts a universal Schwarzschild exterior for both stars and black holes, quantum corrections are sourced only by matter. As such, quantum effects yield non-trivial deformations in matter-supported spacetimes but leave vacuum geometries effectively unchanged \cite{Fabbri:2005nt,Alonso-Bardaji:2020rxb}. This distinction has also been highlighted in broader discussions of the vacuum's role in spacetime structure, suggesting that quantum gravity recognizes a deeper interplay between geometry and energy-momentum content \cite{Overduin:2001uz,Carlip:2009km}.

This discrepancy prompts an important conceptual and observational question: can the gravitational field around compact stars be distinguished from that of black holes through weak-field experiments? Specifically, are there subtle but measurable imprints—such as those embedded in light deflection, gravitational redshift, perihelion precession, or the Shapiro time delay—that could reveal the quantum nature of gravity in stellar environments?

In pursuit of these questions, we focus on a recently proposed quantum-corrected star metric, derived within the EFT approach and characterized by logarithmic non-localities involving the Ricci scalar and tensor. This metric introduces modifications to the $ g_{tt} $ and $ g_{rr} $ components in the exterior region of a star with constant energy density, thereby altering the geodesic motion of both massive and massless test particles. Importantly, these corrections scale as $ \mathcal{O}(G^2M/r^3) $, rendering them subleading in the weak-field regime, yet potentially accessible to ultra-precise measurements.

Gravitational lensing by black holes offers a unique probe of spacetime curvature, with the deflection angle of light rays encoding both the mass and the detailed properties of the lensing object. In the weak field limit, where the impact parameter \(b\) greatly exceeds the Schwarzschild radius \(r_s = 2GM/c^2\), the bending angle admits a perturbative expansion in \(r_s/b\), beginning with the classic result \(\Delta\phi \approx 4GM/(c^2 b)\). Virbhadra et al. first provided closed‐form expressions for image positions and magnifications in this regime for a Schwarzschild lens \cite{Virbhadra:1999nm}, while subsequent work by Virbhadra et al. incorporated scalar‐field effects and higher‐order corrections \cite{Virbhadra:1998dy}. Extensions to charged black holes in the Reissner Nordstrom metric have been constrained observationally, with Zakharov deriving upper bounds on both electric \cite{Zakharov:2014lqa} and tidal charges \cite{Zakharov:2021gbg} from Event Horizon Telescope data. In this paper, we revisit the weak deflection angle for general spherically symmetric black holes, derive analytic expressions up to second post-Newtonian order, and assess their impact on precision lensing observations. Gibbons and Werner introduce a novel method to compute the weak deflection angle by applying the Gauss-Bonnet theorem to the optical geometry of a lensing spacetime \cite{Gibbons:2008rj}.  Li, Zhang, and \"Ovg\"un demonstrate that the presence of circular particle orbits around a compact object introduces analytic corrections to the weak gravitational lensing deflection angle \cite{Li:2020wvn}. It is proved the equivalence between the Gibbons-Werner surface integral approach and the traditional null geodesic method for calculating light deflection in weak fields \cite{Li:2019mqw}. One employs a Jacobi metric framework to derive the finite-distance gravitational deflection of massive particles, highlighting corrections beyond the infinite distance approximation \cite{Li:2019qyb}. The gravitational lensing effect of black holes in effective quantum gravity exhibits quantum-induced deviations from classical deflection angles, indicating observable imprints of quantum corrections \cite{Liu:2024wal}. Kudo and Asada establish a correspondence between two commonly used gravitational lens equations in static, spherically symmetric spacetimes, thereby unifying the treatment of weak deflection regimes \cite{Kudo:2024aak}. Takizawa and Asada apply a constant-curvature optical background formalism to Weyl gravity, deriving modified weak deflection angle expressions for lensing in this alternative theory \cite{Takizawa:2023izb}. Ishihara et al. generalize the finite-distance light bending angle by incorporating both source and observer locations via the Gauss-Bonnet theorem in optical geometry \cite{Ishihara:2016vdc}. Virbhadra and Keeton compute the time delay and centroid shift of images formed by black holes and naked singularities, elucidating weak field lensing observables \cite{Virbhadra:2007kw}. Keeton and Petters develop a post-Newtonian formalism to test theories of gravity using lensing by static, spherically symmetric compact objects in the weak deflection limit \cite{Keeton:2005jd}.  Kochanek, Keeton, and McLeod emphasize the diagnostic power of Einstein rings as precise probes of lens mass distributions in weak gravitational lensing surveys \cite{Kochanek:2000ue}.  Fathi et al. investigate how a Chaplygin Jacobi dark fluid surrounding a black hole alters its shadow and weak deflection angle, quantifying observable deviations \cite{Fathi:2024kda}. Particle dynamics and gravitational weak lensing effects in Kalb-Ramond gravity have been studied, revealing significant deviations from classical results \cite{Atamurotov:2022wsr}. Weak gravitational lensing by a naked singularity surrounded by plasma exhibits unique signatures distinguishable from standard black hole scenarios \cite{Atamurotov:2022srw}. Gravitational weak lensing by a black hole in Horndeski gravity, considering plasma effects, demonstrates distinct observational characteristics compared to General Relativity \cite{Atamurotov:2022slw}. The shadow of a rotating non-Kerr black hole shows clear differences in shape and size from the Kerr black hole shadow, providing a potential test for alternative gravity theories \cite{Atamurotov:2013sca}. Plasma presence around black holes significantly alters their optical appearance, affecting the size and shape of the shadow observable by distant observers \cite{Atamurotov:2015nra}. Optical properties, including gravitational lensing and retrolensing, of braneworld black holes differ notably from their classical counterparts, opening avenues for observational tests of braneworld cosmology \cite{Abdujabbarov:2017pfw}. Quantum corrections within effective field theories alter both the dynamical oscillation spectra and the shadow silhouette of Schwarzschild black holes, revealing signatures of underlying quantum gravity effects \cite{Wang:2025fmz}. Araujo Filho et al. explore the influence of an antisymmetric tensor field on charged black hole lensing, including its effects on deflection angles, scattering, and QNMs \cite{AraujoFilho:2024rcr}.  The weak gravitational lensing signatures of Lorentz-violating black holes are analyzed, revealing characteristic deviations from GR predictions \cite{Filho:2024isd}. Afrin et al. use the Event Horizon Telescope observations of Sgr A* to place novel constraints on loop quantum gravity-inspired modifications of the black hole spacetime \cite{Afrin:2022ztr}. Calza et al. construct non-time-radial-symmetric and loop quantum gravity–inspired metrics for primordial regular black holes, demonstrating their viability as all of the dark matter \cite{Calza:2024xdh}.  Vagnozzi et al. perform horizon-scale tests of gravity theories and fundamental physics by analyzing the Event Horizon Telescope image of Sagittarius A*, deriving limits on deviations from GR \cite{Vagnozzi:2022moj}. Strong gravitational lensing and black hole shadow observations can place stringent constraints on modified gravity theories, effectively bounding deviations from General Relativity in the strong field regime \cite{Kuang:2022ojj}. Analyses of photon regions and shadow observables for a charged rotating black hole, calibrated against M87* data, provide limits on the black hole’s charge and spin parameters \cite{Meng:2022kjs}. Revisiting black hole shadows and weak gravitational lensing with Chern-Simons modifications uncovers measurable departures in deflection angles and shadow morphology due to the parity-violating term \cite{Meng:2023wgi}. The study of deflection angles and shadow behaviors of quintessential black holes in arbitrary dimensions demonstrates how both the quintessence parameter and spacetime dimensionality jointly influence light bending and apparent shadow size \cite{Belhaj:2020rdb}.  
 
Quasinormal modes (QNMs) describe characteristic oscillations of perturbations around black hole spacetimes, encoding valuable information about the underlying gravitational geometry and dynamics. These modes are complex frequencies whose real parts represent oscillation frequencies, while the imaginary parts signify damping rates associated with the dissipation of energy into the event horizon or radiation at infinity \cite{Berti:2009kk, Konoplya:2011qq,Cardoso:2001hn,Cardoso:2001bb}. Determining QNMs analytically is often challenging, particularly for complicated black hole solutions or higher-dimensional spacetimes; hence, numerical and semi-analytical methods become indispensable tools. The Wentzel-Kramers-Brillouin (WKB) approximation has emerged as one of the most widely used and efficient semi-analytic techniques for estimating QNMs, especially in scenarios where potential barriers dominate wave propagation near black holes \cite{Konoplya:2003ii, Konoplya:2019hlu}. By systematically improving its accuracy through higher-order corrections, the WKB method provides a practical and reliable approach to computing QNM spectra for a variety of perturbation problems \cite{Cardoso:2008bp, Konoplya:2019hlu}.   Dolan analyzed instabilities of massive Klein-Gordon fields on the Kerr spacetime, highlighting conditions leading to unstable modes \cite{Dolan:2007mj}. Cardoso et al. connected geodesic stability and Lyapunov exponents with quasinormal mode calculations for black holes \cite{Cardoso:2008bp}. Berti, Cardoso, and Starinets reviewed QNMs of black holes and black branes, covering various theoretical and astrophysical contexts \cite{Berti:2009kk}. Konoplya and Zhidenko extensively reviewed QNMs in diverse physical scenarios from astrophysics to string theory \cite{Konoplya:2011qq}. Pani et al. examined perturbations of slowly rotating Kerr black holes, focusing on massive vector fields \cite{Pani:2012bp}. Cardoso, Franzin, and Pani questioned whether gravitational-wave ringdown signals can serve as direct probes of the black hole event horizon \cite{Cardoso:2016rao}. Konoplya, Zhidenko, and Zinhailo presented a higher-order WKB formula for efficiently calculating QNMs and grey-body factors \cite{Konoplya:2019hlu}. The QNMs and absorption cross-section of a Bardeen black hole surrounded by perfect fluid dark matter in four dimensions were computed, highlighting the influence of dark matter on black hole oscillations \cite{Rincon:2025buq}. Quasinormal modes of scale-dependent black holes in (1+2)-dimensional Einstein-power-Maxwell theory reveal sensitivity to scale-dependent couplings, providing insights into lower-dimensional gravity models \cite{Rincon:2018sgd}. Greybody factors and QNMs for nonminimally coupled scalar fields in a cloud of strings within a (2+1)-dimensional background indicate notable effects due to scalar field coupling and string cloud parameters \cite{Rincon:2018ktz}. Analysis of QNMs of an improved Schwarzschild black hole demonstrates significant corrections to classical predictions, offering possible observational implications for quantum gravity phenomenology \cite{Rincon:2020iwy}. Quasinormal modes of black holes with non-linear electrodynamic sources in Rastall gravity exhibit distinct oscillation spectra influenced by both the nonlinear fields and the modified conservation law \cite{Gogoi:2021dkr}. Constraints on QNMs derived from black hole shadows in regular non-minimal Einstein-Yang-Mills gravity provide observational limits linking shadow size to the spectrum of ringdown frequencies \cite{Gogoi:2024vcx}. The QNMs and greybody factors of de Sitter black holes surrounded by quintessence in Rastall gravity reveal that the presence of quintessence significantly alters both the damping rates and transmission probabilities of perturbations \cite{Gogoi:2023lvw}. The shadow and QNMs of rotating Einstein-Euler-Heisenberg black holes show that nonlinear electrodynamic corrections produce measurable shifts in the shadow boundary and the oscillation frequencies of the ringdown signal \cite{Lambiase:2024lvo}. Loop quantum gravity effects introduce distinctive anomalies in quasinormal mode spectra, such as unexpected splitting and shifts in damping rates, pointing to underlying quantum corrections \cite{Fu:2023drp}. Quasinormal modes and Hawking radiation spectra of a charged Weyl black hole reveal characteristic imprints of the Weyl charge on both oscillation frequencies and particle emission rates \cite{Fu:2022cul}. Scalar field perturbations around a rotating hairy black hole exhibit unique quasinormal mode families, quasibound states, and superradiant instabilities tied to the hair parameters \cite{Lei:2023wlt}. Analysis of QNMs for high-dimensional RN-AdS black holes across phase transitions reveals that the dynamic perturbation spectrum mirrors thermodynamic behavior, underscoring a deep connection between stability properties and black hole phase structure \cite{Chabab:2016cem}. 
 
 To probe the empirical consequences of such quantum corrections, we examine their effects on canonical weak field observables. First, we employ the Gauss-Bonnet theorem to compute the deflection angle of light and time-like geodesics, explicitly tracking the dependence on the EFT coefficients $ \hat{\alpha} $ and $ \hat{\beta} $. Next, we calculate corrections to the Shapiro time delay and gravitational redshift, both of which are sensitive to modifications in the temporal metric component. Finally, we analyze the perihelion advance of planetary orbits in this background and extract the quantum contribution to the precession rate.

By comparing these results with their classical general relativistic counterparts, we aim to elucidate the observational viability of detecting quantum gravitational signatures in weak-field astrophysical tests. Although the magnitudes of the corrections are exceedingly small, they provide an unambiguous theoretical signal distinguishing quantum-corrected stellar spacetimes from classical black holes. Moreover, these findings may be especially relevant in regimes where quantum gravity effects are non-negligible, such as in Planck-scale compact objects or early-universe cosmology. In this regard, our study contributes to the growing body of evidence that effective quantum gravity leaves faint but discernible imprints even in the ostensibly classical domain of weak gravitational fields.

The paper is organized as follows: In Sec. \ref{sec2}, we review the quantum-corrected star metric and identify the leading-order modifications to the spacetime geometry. Sec. \ref{sec3} presents our derivation of the weak deflection angle via the Gauss-Bonnet theorem, while the succeeding subsections focus on corrections to the Shapiro delay and gravitational redshift, and analysis of the perihelion advance of a planetary orbit. In Sec. \ref{perturbations}, we study scalar waves propagation and QNMs. Finally, Sec. \ref{conc} concludes with a discussion of our results and their broader implications for testing quantum gravity. We use the metric signature $(-,+,+,+)$ and geometric units $\hbar=c=1$ throughout the analysis in the paper.

\section{Star-like metric with quantum-correction} \label{sec2}
\subsection{Brief review}
The quest for a consistent quantum theory of gravity remains one of the most formidable challenges in modern theoretical physics. However, a fruitful and operational approach has emerged in the form of EFT, where quantum gravitational corrections are systematically computed at energy scales well below the Planck scale, $ M_{\text{Pl}} = \sqrt{\hbar c/G} $. In this low-energy regime, quantum corrections manifest as higher-derivative terms in the gravitational action, with both local and non-local structures governed by general covariance and the particle content of the underlying theory.

Calmet et al. \cite{Calmet:2019eof} perform a detailed computation of quantum gravitational corrections to the spacetime metric generated by a static, spherically symmetric star in equilibrium. The starting point is the effective action of the form \cite{Barvinsky:1983vpp,Barvinsky:1985an,Barvinsky:1987uw,Donoghue:1994dn,Calmet:2017qqa,Calmet:2018elv}
\begin{widetext}
\begin{align} \label{action}
\Gamma[g] &= \int d^4x \sqrt{-g} \left[ \frac{R}{16\pi G_N} + c_1 R^2 + c_2 R_{\mu\nu} R^{\mu\nu} + c_3 R_{\mu\nu\rho\sigma} R^{\mu\nu\rho\sigma} \right]  \nonumber \\
&-\int d^4x \sqrt{-g} \left[ \alpha R \ln \left( \frac{\Box}{\mu^2} \right) R + \beta R_{\mu\nu} \ln \left( \frac{\Box}{\mu^2} \right) R^{\mu\nu} + \gamma R_{\mu\nu\rho\sigma} \ln \left( \frac{\Box}{\mu^2} \right) R^{\mu\nu\rho\sigma} \right],
\end{align}
\end{widetext}
where the non-local logarithmic terms originate from one-loop quantum corrections due to massless fields (scalars, fermions, vectors, and gravitons). The coefficients $ \alpha, \beta, \gamma $ are universal for a given particle content and do not depend on the ultraviolet completion of quantum gravity, making their predictions model-independent within the EFT regime. Thought for a few seconds

The quadratic-curvature extension of Einstein gravity propagates three mass eigenstates: the usual massless spin-2 graviton, a massive spin-2 excitation, and a massive scalar (spin-0) excitation.  Of these, only the massive spin-2 mode carries the wrong-sign (ghost) kinetic term, as shown in Refs. \cite{Calmet:2018hfb,Calmet:2018uub}, however, it does not introduce any physical inconsistency at the classical level.  Indeed, once one integrates out the graviton fluctuations, the resulting effective action can be viewed as containing classical fields corresponding to these massive modes.  In particular, the massive spin-2 field appears with an overall negative coupling to the stress-energy tensor, so that it mediates a repulsive force.

In their paper, they consider a compact star with constant energy density $ \rho(r) = \rho_0 \, \Theta(R_s - r) $, where $ R_s $ is the stellar radius. The classical background metric consists of the standard Schwarzschild interior solution for $ r < R_s $ and the Schwarzschild vacuum solution for $ r > R_s $. Quantum corrections $ g_{\mu\nu} \mapsto g_{\mu\nu} + h_{\mu\nu} $ are computed perturbatively to leading nontrivial order $ \mathcal{O}(G_N^2) $. Importantly, while the Schwarzschild exterior has vanishing Ricci tensor and scalar (and hence receives no corrections at this order in the vacuum), the interior solution, supported by nonzero stress-energy, gives rise to non-trivial quantum corrections even outside the star.

%The corrected components of the metric are given for $ r > R_s $ as:\begin{equation}\delta g_{tt}^{\text{ext}}(r) = \frac{G_N^2 M}{R_s^3} ( \alpha + \beta + 3\gamma ) \left[ \frac{2 R_s}{r} + \ln\left( \frac{r - R_s}{r + R_s} \right) \right],\end{equation}\begin{equation}\delta g_{rr}^{\text{ext}}(r) = \frac{G_N^2 M}{r(r^2 - R_s^2)} ( \alpha - \gamma ),\end{equation}where $ M = \frac{4\pi}{3} \rho_0 R_s^3 $ is the Misner-Sharp mass of the source. These expressions illustrate that the corrections to Newton's potential are of order $ \mathcal{O}(G_N^2) $, and vanish asymptotically, ensuring the ADM mass remains $ M $. In the harmonic gauge, these corrections enter the weak-field expansion as additional terms $ \sim 1/r^3 $ in the metric potentials.

The distinction between star and black hole geometries becomes profound when one considers the black hole limit. Eternal Schwarzschild black holes, being vacuum solutions, are unaffected by these non-local terms to second order in curvature. Yet, the quantum-corrected star metric retains corrections for all finite radii, including near the Buchdahl limit $ R_s > \frac{9}{8} R_H = \frac{9}{4} G_N M $. This discontinuity poses a question of fundamental importance: if a star collapses to form a black hole, what becomes of the quantum corrections? Are they abruptly "turned off" as soon as a horizon forms? If so, how does this reconcile with the expected continuity of the gravitational field in semiclassical gravity?

Pursuing this line of thought, they conjectured that quantum black holes, particularly those near the Planck scale $ M \sim M_{\text{Pl}} $, cannot be meaningfully described by vacuum metrics. In such cases, the interior of the object is expected to harbor extreme curvature fluctuations where classical notions of geometry break down. The quantum-corrected star metric, despite being static, may offer a better approximation to the gravitational field around such objects. Notably, they derive that the gravitational radius $ R_H $, defined implicitly via the corrected $ g_{rr}^{-1}(R_H) = 0 $, can be larger than the Schwarzschild radius, hinting at the modification of horizon structure due to quantum effects.

The mathematical subtleties of non-local operators acting on discontinuous matter profiles are also addressed. The divergence at $ r = R_s $ is traced to the Heaviside step function in $ \rho(r) $, which is not in the domain of the operator $ \ln(\Box) $. A regularization is proposed by smoothing the density profile over a finite width $ \epsilon $, ensuring that physical divergences are avoided and the non-local corrections remain under control.

As for the substantial implications, observationally and in principle, one could distinguish a collapsing star from an eternal black hole by detecting corrections to the gravitational potential at order $ G_N^2 $—a prospect extremely challenging in practice but not ruled out in principle. Conceptually, the results imply that quantum gravity breaks the classical degeneracy between matter-supported and vacuum solutions. This could bear on how semiclassical black holes are modeled, particularly in early-universe cosmology or high-energy regimes. Lastly, the results stress the importance of metric regularity and well-defined curvature operators in theories with non-local corrections. This may also connect with the broader program of quantum gravitational effective actions and their role in non-perturbative collapse or horizon avoidance mechanisms. It only reaffirms that even within the well-trodden domain of general relativistic solutions, quantum gravity introduces distinctions that challenge classical intuition. The implications for black hole thermodynamics, horizon formation, and quantum geometry remain rich and invite deeper analysis through both analytic and numerical approaches.

\subsection{The Quantum-Corrected Exterior Metric}
In the EFT approach to quantum gravity, the classical Schwarzschild solution describing the exterior spacetime of a spherically symmetric, static matter distribution is modified by non-local, curvature-squared corrections at order $ \mathcal{O}(G_N^2) $. These corrections arise from the effective gravitational action expanded around a classical background and integrated over fluctuations of the massless fields, such as scalars, fermions, gauge bosons, and the graviton, under the assumption of general coordinate invariance.

For the case of a homogeneous, isotropic stellar object with radius $ R_s $ and total mass $ M $, the quantum-corrected line element in the region $ r > R_s $ takes the form \cite{Calmet:2019eof}
\begin{equation}
ds^2 = -\Bigl[1 - \frac{2GM}{r}
       - \frac{\hat{\alpha}G^2M}{r^3}
       + \mathcal{O}(G^3)\Bigr]\,dt^2 + \Bigl[\bigl(1 - \tfrac{2GM}{r}\bigr)^{-1}
       + \frac{\hat{\beta}G^2M}{r^3}
       + \mathcal{O}(G^3)\Bigr]\,dr^2
       + r^2\,d\Omega^2 \,,
\label{metric1}
\end{equation}
where $ d\Omega^2 \equiv d\theta^2 + \sin^2\theta \, d\phi^2 $, and $ G \equiv G_N $ is Newton's gravitational constant. The coefficients $ \hat{\alpha} $ and $ \hat{\beta} $ encapsulate the quantum corrections to the metric and are constructed from combinations of the universal Wilson coefficients $ \alpha, \beta, \gamma $, appearing in the non-local terms of the effective action (second term of Eq. \eqref{action}).
Specifically, the combinations relevant to the corrections are (see Table 1 of Ref. \cite{Calmet:2019eof}, using the graviton loop value):
\begin{equation} \label{graviton}
    \hat{\alpha} \equiv 128\pi (\alpha + \beta + 3\gamma) \sim 4.52, \qquad \hat{\beta} \equiv 384\pi (\alpha - \gamma) \sim -1.85.
\end{equation}

From here on, we write $ds^2$ as a $1+2$ dimensionality metric by specializing analysis at the equatorial plane: $\theta = \pi/2$:
\begin{equation} \label{metric2}
    ds^2 = -A(r)dt^2 + B(r) dr^2 + C(r) d\phi^2.
\end{equation}

These parameters reflect the cumulative contributions of loop-induced vacuum polarization from all massless species. Each field type contributes distinctively: scalar fields depend on their non-minimal coupling $ \xi $, while vector, fermionic, and gravitational degrees of freedom yield fixed contributions. Their values are gauge-independent when derived through a properly regulated background field method, ensuring the observability of the resulting physical effects.

The $A$ component of the metric governs gravitational redshift, energy levels, and photon trajectories. Its quantum correction via $ \hat{\alpha} $ introduces a deviation from Newton's potential of the form $ \sim G^2 M/r^3 $. This modifies the effective potential and, consequently, impacts weak-field observables such as the deflection of light rays grazing the stellar limb. Meanwhile, the radial component $ g_{rr} $, corrected by $ \hat{\beta} $, affects the geometry of spatial slices and thus modifies geodesic curvature, with implications for perihelion precession and radar time delays.

The corrections fall off rapidly with radial distance, and their contributions are subleading in the solar system context. Nevertheless, they encode essential quantum gravitational signatures that, in principle, allow for the operational distinction between matter-supported spacetimes and classical vacuum solutions. From a phenomenological standpoint, the key parameters to constrain are $ \hat{\alpha} $ and $ \hat{\beta} $, given their direct role in modifying weak-field observables. Constraints may be obtained through precision tests of GR in the solar system, including Mercury's perihelion precession, gravitational lensing, and Shapiro delay.

Notably, while the classical exterior Schwarzschild metric applies universally to both stars and black holes, these quantum corrections break this degeneracy. Static vacuum solutions, such as eternal black holes, do not receive corrections at this order, whereas stellar spacetimes do. This suggests that the parameters $ \hat{\alpha} $ and $ \hat{\beta} $ may also play a critical role in modeling the spacetime geometry of Planck-scale black holes, for which the vacuum approximation is likely to break down due to non-perturbative quantum effects.

\section{Solar System Tests} \label{sec3}
\subsection{Weak deflection angle}
In this section, we calculate the deflection angle of a massive particle (represented by $\mu = 1$) grazing on a star with quantum gravitational correction. In other words, our goal is to obtain a general expression for the weak deflection angle, which describes the deflection of both time-like and null particles.

With the metric in Eq. \eqref{metric2}, the Jacobi metric, reads
\begin{equation}
    dl^2=g_{ij}dx^{i}dx^{j} =[E^2-\mu^2A(r)]\left(\frac{B(r)}{A(r)}dr^2+\frac{C(r)}{A(r)}d\Omega^2\right),
\end{equation}
where $d\Omega^2=d\theta^2+\sin^2\theta d\phi^2$ is the line element of the unit two-spheres. Here, $E$ is the energy of the massive particle defined by
\begin{equation} \label{en}
    E = \frac{1}{\sqrt{1-v^2}},
\end{equation}
where $v$ is the particle's velocity. Without loss of generality, as we specialize along the equatorial plane ($\theta = \pi/2$), the Jacobi metric is simplified to
\begin{equation} \label{eJac}
    dl^2=(E^2-A(r))\left(\frac{B(r)}{A(r)}dr^2+\frac{C(r)}{A(r)}d\phi^2\right)
\end{equation}
The determinant of the Jacobi metric above can also be easily calculated as
\begin{equation}
    g=\frac{B(r)C(r)}{A(r)^2}(E^2-A(r))^2.
\end{equation}
Next, we will use these equations to find the weak deflection angle using the Gauss-Bonnet theorem (GBT) \cite{Gibbons:2008rj}:\\
\textbf{Theorem:} \textit{Let $D$ be a smooth, compact, freely orientable surface with boundary $ \partial M $, embedded in $\mathbb{R}^3$. Let $K$ be the Gaussian curvature of $D$ at each point, $k_g$ is the geodesic curvature along the boundary, and $ds$ represents the line element along the boundary of $D$. Then, the GBT can be written as}
\begin{equation} \label{eGBT}
    \iint_DKdS+\sum\limits_{a=1}^N \int_{\partial D_{a}} \kappa_{\text{g}} d\ell+ \sum\limits_{a=1}^N \theta_{a} = 2\pi\chi(D).
\end{equation}
Since $D$ is a non-singular region, its Euler characteristic is $\chi(D) = 1$. Lastly, $\theta_\text{a}$ is the jump angle. Eq. \eqref{eGBT} has been used widely in the literature to probe the effect of certain parameters, like quantum corrections from different black hole models, on gravitational lensing, especially in the weak regime \cite{Alloqulov:2024xak,Alloqulov:2025htt,Ghaffarnejad:2015jka,Ditta:2025vsa,Pantig:2024fbh,Alonso-Bardaji:2020rxb,Junior:2023xgl,Ditta:2025ezx,Lobos:2022jsz,Lambiase:2024vkz,Pantig:2024asu}. 

The GBT is so useful when it comes to asymptotically flat spacetimes. However, it becomes problematic when it encounters non-asymptotically flat spacetimes. For such a case, it was shown by \cite{Li:2020wvn} that Eq. \eqref{eGBT} can be written as
\begin{equation} \label{ewda_li}
    \Theta = \int^{\phi_{\rm R}}_{\phi_{\rm S}} \int_{r_{\rm ph}}^{r(\phi)} K dS + \phi_{\rm RS}.
\end{equation}
Here, the separation angle between the source and the receiver is $\phi_{\rm RS} = \phi_{\rm R}-\phi_{\rm S}$, where $\phi_{\rm R} = \pi - \phi_{\rm S}$. The infinitesimal curve surface $dS$ is also given by
\begin{equation}
    dS = \sqrt{g}drd\phi.
\end{equation}
The first step of deriving the general expression for the weak deflection angle is to find the orbit equation:
\begin{eqnarray}
F(u) = \left(\frac{du}{d\phi}\right)^2 
\label{Fu_split}= \frac{C(u)^2\,u^4}{A(u)\,B(u)}
   \Biggl[
     \Bigl(\tfrac{E}{J}\Bigr)^2
     - A(u)\Bigl(\tfrac{1}{J^2}+\tfrac{1}{C(u)}\Bigr)
   \Biggr]\,.
\end{eqnarray}
where we have used the substitution $r = 1/u$ and $J = v b E$ is the angular momentum of the massive particle, and $b$ is the impact parameter.

With the metric coefficients, we find the orbit equation in terms of $u$ as
\begin{eqnarray}\label{e_time-orb}
F(u) = \frac{E^2 - 1}{J^2} - u^2
  + 2\,u \Bigl(u^2 + \tfrac{1}{J^2}\Bigr)\,GM
- u^3 \Bigl(\tfrac{E^2}{J^2} - \tfrac{1}{J^2} - u^2\Bigr)\,\hat{\beta}\,G^2M+ u^3 \Bigl(\tfrac{1}{J^2} + u^2\Bigr)\,\hat{\alpha}\,G^2M.
\end{eqnarray}
For time-like orbit, an excellent guess solution to the Schwarzschild case is the expression
\begin{equation} \label{euphi}
    u(\phi) = \frac{\sin(\phi)}{b}+\frac{1+v^2\cos^2(\phi)}{b^2v^2}M.
\end{equation}
Instead of solving the difficult differential equation shown in Eq. \eqref{e_time-orb} that contains the quantum correction parameter, we can resort to iterative methods by adding a term to Eq. \eqref{euphi} with $p \hat{\alpha} G^2 M$. The goal is to solve $p$ by iteration. After doing so, we found that $p=0$ for the iteration in $\hat{\alpha}$ and $\hat{\beta}$, hence, it seems that the quantum correction does not affect $u(\phi)$.

The Gaussian curvature $K$, in terms of affine connections and determinant $g$, is defined as
\begin{align}
K= \frac{1}{\sqrt{g}}
   \Biggl[
     \frac{\partial}{\partial\phi}\Bigl(\frac{\sqrt{g}}{g_{rr}}\Gamma_{rr}^{\phi}\Bigr)
     - \frac{\partial}{\partial r}\Bigl(\frac{\sqrt{g}}{g_{rr}}\Gamma_{r\phi}^{\phi}\Bigr)
   \Biggr]= -\,\frac{1}{\sqrt{g}}
   \frac{\partial}{\partial r}
   \Bigl(\frac{\sqrt{g}}{g_{rr}}\Gamma_{r\phi}^{\phi}\Bigr)\,.
\label{K_simplified}
\end{align}

since $\Gamma_{rr}^{\phi} = 0$ for Eq. \eqref{eJac}. Then with the analytical solution to $r_\text{ph}$,
\begin{equation}
    \left[\int K\sqrt{g}dr\right]\bigg|_{r=r_\text{ph}} = 0,
\end{equation}
thus,
\begin{align}
\int_{r_{\rm ph}}^{r(\phi)} &K\sqrt{g}\,dr 
\label{gct_split}
= -\frac{1}{2}
   \Biggl[
     \frac{A(r)\bigl(E^2 - A(r)\bigr)\,C'(r)
           - E^2\,C(r)\,A'(r)}
          {A(r)\bigl(E^2 - A(r)\bigr)\,\sqrt{B(r)\,C(r)}}
   \Biggr]_{r = r(\phi)}\,. 
%\nonumber
\end{align}

The prime denotes differentiation with respect to $r$. 
\begin{align}\label{gct_expanded}\
\int_{r_{\rm ph}}^{r(\phi)} K\sqrt{g}\,dr 
&\sim -1 
  + \frac{\sin(\phi)\,(2E^2 - 1)\,GM}{(E^2 - 1)\,b}
  + \frac{3\,\sin^3\!\phi \,E^2\,\hat\alpha\,G^2M}
         {b^3\,(2E^2 - 2)}
  + \frac{\sin^3\!\phi \,\hat\beta\,G^2M}{2\,b^3}
  + \hat{C} + \mathcal{O}(M^2,\hat{\alpha}^2,\hat{\beta}^2)\,.
\end{align}
where $\hat{C}$ is an integration constant.
We want to express the next integration as an indefinite integral to see the overall structure of the equation. The result is given as
\begin{eqnarray}\label{gauss}
&
\iint_{r_{\rm ph}}^{r(\phi)} K\sqrt{g}\,dr\,d\phi 
\sim -\phi 
  - \frac{\cos(\phi)\,(2E^2 - 1)\,GM}{(E^2 - 1)\,b} - \frac{E^2\,\hat{\alpha}\,G^2M\,(2 + \sin^2\!\phi)\,\cos(\phi)}
             {2\,b^3\,(E^2 - 1)} - \frac{\hat{\beta}\,G^2M\,(2 + \sin^2\!\phi)\,\cos(\phi)}
             {6\,b^3}
  + \hat{C} + \mathcal{O}(M^2,\hat{\alpha}^2,\hat{\beta}^2)\,.
\end{eqnarray}

Thus, using Eq. \eqref{ewda_li}, we find the initial expression for the weak deflection angle as
\begin{eqnarray} \label{ewda}
\Theta & \sim\; &\frac{2\cos\!\phi\,(2E^2 - 1)\,GM}{(E^2 - 1)\,b} + \frac{E^2\,\hat{\alpha}\,G^2M\,(2 + \sin^2\!\phi)\,\cos\!\phi}
        {b^3\,(E^2 - 1)}+ \frac{\hat{\beta}\,G^2M\,(2 + \sin^2\!\phi)\,\cos\!\phi}
        {3\,b^3}
 + \mathcal{O}(M^2,\hat{\alpha}^2,\hat{\beta}^2)\,.
\end{eqnarray}
The next objective is to obtain the expression for $\phi$ through Eq. \eqref{euphi},
\begin{equation}
    \phi = \arcsin(bu)+\frac{GM\left[v^{2}\left(b^{2}u^{2}-1\right)-1\right]}{bv^{2}\sqrt{1-b^{2}u^{2}}}\,.
\end{equation}
Using the following properties
\begin{align}
    \cos \left(\pi - \phi\right) = -\cos \left(\phi\right), \qquad \sin^2 \left(\pi - \phi\right) = \sin^2 \left(\phi\right),
\end{align}
we obtain
\begin{align} \label{cs}
    \cos(\phi) \sim \sqrt{1-b^{2}u^{2}}-\frac{GMu\left[v^{2}\left(b^{2}u^{2}-1\right)-1\right]}{v^{2}\sqrt{1-b^{2}u^{2}}} + \mathcal{O}(M^2),
\end{align}
and
\begin{align} \label{cs2}
    \sin^2(\phi) \sim b^{2} u^{2}+\frac{u \left[-1+\left(b^{2} u^{2}-1\right) v^{2}\right] GM}{v^{2}} + \mathcal{O}(M^2).
\end{align}

From Eq. \eqref{en} and by plugging Eqs. \eqref{cs}-\eqref{cs2} in Eq. \eqref{ewda}, we finally get
\begin{align}\label{theta_split} 
\Theta 
&\sim \frac{2GM\,(v^{2}+1)}{b\,v^{2}}\sqrt{1 - b^{2}u^{2}}
+ \frac{\hat{\alpha}\,G^2M}{v^{2}b^{3}}
   \Bigl[(b^{2}u^{2}+2)\sqrt{1 - b^{2}u^{2}}\Bigr] + \frac{\hat{\beta}\,G^2M}{3\,b^{3}}
   \Bigl[(b^{2}u^{2}+2)\sqrt{1 - b^{2}u^{2}}\Bigr]
   + \mathcal{O}(M^2,\hat{\alpha}^2,\hat{\beta}^2)\,.
\nonumber
\end{align}
The above is general as the finite distance of the source and the receiver (assumed to be equidistant) from the black hole is considered. When $u$ approaches zero, then the above equation reduces to
\begin{equation} \label{ewda_time}
    \Theta^{\rm timelike} \sim \frac{2 \left(v^{2}+1\right) GM}{v^{2} b}+\frac{2 \hat{\alpha} G^2 M}{v^{2} b^{3}}+\frac{2 \hat{\beta} G^2 M}{3 b^{3}} + \mathcal{O}(M^2,\hat{\alpha}^2,\hat{\beta}^2).
\end{equation}
For the weak deflection of photons ($v = 1$),
\begin{equation} \label{ewda_null}
    \Theta^{\rm null} \sim \frac{4 GM}{b}+\frac{2 \hat{\alpha} G^2 M}{b^{3}}+\frac{2 \hat{\beta} G^2 M}{3 b^{3}} + \mathcal{O}(M^2,\hat{\alpha}^2,\hat{\beta}^2).
\end{equation}
As we see, the weak deflection angle is sensitive to the quantum corrections $\hat{\alpha}$ and $\hat{\beta}$. It is easy to see how this will reduce to the Schwarzschild case when there are no quantum corrections. The classical GR prediction of $1.75$ arcsecond is expected on the first term. For the quantum correction contribution, it is only approximately $7.043 \times 10^{-18}$ arcsecond. In obtaining these values, we set $G = 1$, and used solar parameters such as $M_{\odot} = 1477 \text{ m}$, and $b = R_{\odot} = 6.9634 \times 10^{8} m$.

\subsection{Perihelion Advance of Mercury}
Among the classical tests of GR, the anomalous perihelion precession of planetary orbits, most notably Mercury's, provided one of the earliest confirmations of Einstein's theory \cite{Einstein:1915bz}. In this section, we derive the correction to the perihelion precession angle per orbit due to quantum gravitational effects, using the EFT corrected metric for a static, spherically symmetric mass distribution. The analysis proceeds by deriving a modified orbit equation under the assumption of test particle motion in a central potential.

We consider a massive test particle (planet) of negligible mass compared to the central body (e.g., Sun), moving in the equatorial plane $ \theta = \pi/2 $ under the corrected spacetime geometry. Using Eq. \eqref{metric2}, we adopt the standard approach of exploiting the spacetime symmetries to reduce the geodesic equations. Due to time-translational and rotational invariance, we define the conserved quantities:
\begin{equation} \label{e_cons}
    E \equiv -A(r) \frac{dt}{d\tau}, \qquad J \equiv C(r) \frac{d\phi}{d\tau},
\end{equation}
representing the conserved specific energy and angular momentum per unit mass. The normalization condition for the four-velocity of a massive particle $ u^\mu u_\mu = -1 $ yields
%\begin{equation} \label{metric3}
%    -1 = -A(r) \left( \frac{dt}{d\tau} \right)^2 + B(r) \left( \frac{dr}{d\tau} \right)^2 + C(r) \left( \frac{d\phi}{d\tau} \right)^2.
%\end{equation}
%Substituting Eq. \eqref{e_cons} to Eq. \eqref{metric3} leads to
\begin{equation}
    -1 = -\frac{E^2}{A(r)} + B(r) \left( \frac{dr}{d\tau} \right)^2 + \frac{J^2}{C(r)}.
\end{equation}
With a slight rearrangement to solve the radial component,
\begin{equation}
    \left( \frac{dr}{d\tau} \right)^2 = \frac{1}{B(r)}\left( \frac{E^2}{A(r)} - \frac{J^2}{C(r)} - 1\right)
\end{equation}

We now transform the radial coordinate to $ u \equiv 1/r $, and convert derivatives:
\begin{equation}
    \frac{dr}{d\tau} = \frac{dr}{d\phi} \frac{d\phi}{d\tau} = \frac{dr}{d\phi} \frac{J}{C(r)}.
\end{equation}
Thus, the general radial equation becomes
\begin{equation}
    \left( \frac{du}{d\phi} \right)^2 = \frac{u^4 C(u)^2}{A(u)B(u)} \left[ \frac{E^2}{J^2} - A(u)\left(\frac{1}{C(u)} + \frac{1}{J^2}\right) \right].
\end{equation}
This expression can be differentiated with respect to $ \phi $ to yield the orbit equation. However, a more tractable approach, especially at weak field order, is to derive a second-order differential equation for $ u(\phi) $, perturbatively incorporating the corrections due to the modified metric. We now proceed with this method. We begin with the general Schwarzschild result for the classical orbit equation:
\begin{equation}
    \frac{d^2 u}{d\phi^2} + u = \frac{GM}{J^2} + 3GM u^2.
\end{equation}

In our case, the corrections due to $ \hat{\hat{\alpha}} $ and $ \hat{\hat{\beta}} $ modify the effective potential. Following the methods of standard geodesic perturbation theory (e.g., Weinberg, Misner-Thorne-Wheeler), the corrected orbit equation to $ \mathcal{O}(G^2) $ is:
\begin{equation}
    \frac{d^2 u}{d\phi^2} + u = \frac{GM}{J^2} + 3GM u^2 + \delta_{\text{QG}}(u),
\end{equation}
where $ \delta_{\text{QG}}(u) $ is the quantum correction to the right-hand side, which arises from expanding the full geodesic equation with the corrected metric. The result is
\begin{equation}
    \frac{d^2 u}{d\phi^2} + u = \frac{GM}{J^2} + \Lambda u^2,
\end{equation}
where 
\begin{equation}
    \Lambda = 3GM \left[1 + \frac{G E^2 (\hat{\alpha} - \hat{\beta})}{2 J^2} + \frac{G \hat{\beta}}{2 J^2} \right].
\end{equation}

Next, we apply the method of successive approximations by writing,
\begin{equation}
    u(\phi) = u_0(\phi) + \epsilon u_1(\phi) + \cdots,
\end{equation}
where $\epsilon$ is a bookkeeping parameter to track perturbations (set to 1 at the end), and assume that $u_1$ is small. The zeroth-order (Newtonian) equation is:
\begin{equation}
    \frac{d^2 u_0}{d\phi^2} + u_0 = \frac{GM}{J^2},
\end{equation}
which has a general solution of
\begin{equation}
    u_0(\phi) = \frac{GM}{J^2} \left[1 + e \cos(\phi)\right],
\end{equation}
where $e$ is the orbital eccentricity.
Plugging $u(\phi) = u_0 + \epsilon u_1$ into the full equation, and keeping terms only up to first order in $\epsilon$, one gets
\begin{equation}
    \frac{d^2 (u_0 + \epsilon u_1)}{d\phi^2} + (u_0 + \epsilon u_1) = \frac{GM}{J^2} + \Lambda (u_0 + \epsilon u_1)^2,
\end{equation}
and by using expansion $(u_0 + \epsilon u_1)^2 = u_0^2 + 2\epsilon u_0 u_1 + \mathcal{O}(\epsilon^2)$,
%\begin{equation}
%    (u_0 + \epsilon u_1)^2 = u_0^2 + 2\epsilon u_0 u_1 + \mathcal{O}(\epsilon^2)
%\end{equation}
one infers
\begin{equation}
    \frac{d^2 u_0}{d\phi^2} + \epsilon \frac{d^2 u_1}{d\phi^2} + u_0 + \epsilon u_1 = \frac{GM}{J^2} + \Lambda \left( u_0^2 + 2 \epsilon u_0 u_1 \right).
\end{equation}
After we cancel zeroth-order terms and equate first-order parts, we obtain
\begin{equation}
    \frac{d^2 u_1}{d\phi^2} + u_1 = \Lambda u_0^2.
\end{equation}

We then substitute the zeroth-order solution into the first-order equation leading to
\begin{equation}
    u_0 = \frac{GM}{J^2} (1 + e \cos \phi) \Rightarrow u_0^2 = \left( \frac{GM}{J^2} \right)^2 (1 + 2e \cos \phi + e^2 \cos^2 \phi).
\end{equation}
Hence,
\begin{equation}
    \frac{d^2 u_1}{d\phi^2} + u_1 = \Lambda \left( \frac{GM}{J^2} \right)^2 \left(1 + 2e \cos \phi + e^2 \cos^2 \phi \right).
\end{equation}
After using the identity $\cos^2 \phi = (1 + \cos 2\phi)/2$, we obtain
\begin{equation}
    \frac{d^2 u_1}{d\phi^2} + u_1 = \Lambda \left( \frac{GM}{J^2} \right)^2 \left(1 + 2e \cos \phi + \frac{e^2}{2}(1 + \cos 2\phi) \right),
\end{equation}
which is a linear nonhomogeneous differential equation. The particular solution will have terms involving $\phi \sin \phi$ due to resonance with the driving term $ \cos \phi $. The $\cos \phi$ term leads to secular growth (non-periodic part), which is directly related to the perihelion shift.

In extracting the perihelion shift, we write the first-order solution as
\begin{equation}
    u(\phi) = \frac{GM}{J^2} \left[1 + e \cos \left((1 - \delta)\phi\right)\right],
\end{equation}
where $\delta$ is the perihelion precession per revolution, assumed to be small. After using a Taylor expansion $\cos[(1 - \delta)\phi] \approx \cos(\phi - \delta \phi) = \cos \phi + \delta \phi \sin \phi + \cdots$, it causes a slow rotation of the orbit per revolution. The perihelion shift per revolution is then $\Delta \phi = 2\pi \delta$. Standard perturbation theory tells us that the coefficient of the secular term (resonant with $\cos \phi$) gives
\begin{equation} \label{e_phi1}
    \Delta \phi = \frac{6\pi GM}{a(1 - e^2)} + \frac{3\pi (\hat{\alpha} - \hat{\beta}) G^2 M E^2}{a^2 (1 - e^2)^2} + \frac{3\pi \hat{\beta} G^2 M}{a^2 (1 - e^2)^2}.
\end{equation}

Finally, for bound orbits, we can use the Newtonian approximation
\begin{equation}
    E^2 \approx 1 - \frac{GM}{2a},
\end{equation}
and substituting this into \eqref{e_phi1},
\begin{eqnarray}\label{e_precession_multiline}
\Delta\phi 
\approx \frac{6\pi GM}{a(1 - e^2)}
+ \frac{3\pi(\hat{\alpha} - \hat{\beta})\,G^2M}
             {a^2\,(1 - e^2)^2}
    \Bigl(1 - \tfrac{GM}{2a}\Bigr)+ \frac{3\pi\,\hat{\beta}\,G^2M}
             {a^2\,(1 - e^2)^2}\,.
\nonumber
\end{eqnarray}

Thus, the total perihelion advance per revolution is
\begin{eqnarray}
\Delta\phi = \frac{6\pi GM}{a(1 - e^2)}
\label{e_precession_final}+ \frac{3\pi(\hat{\alpha} - \hat{\beta})\,G^2M}
             {a^2\,(1 - e^2)^2}
    \Bigl(1 - \tfrac{GM}{2a}\Bigr)+ \frac{3\pi\,\hat{\beta}\,G^2M}
             {a^2\,(1 - e^2)^2}\,.
\end{eqnarray}
which generalizes the standard GR correction by including terms parametrized by $\hat{\alpha}$ and $\hat{\beta}$, potentially arising from Lorentz-violating or higher-derivative corrections to the gravitational field equations.

We now provide a quantitative analysis of the corrected expression for the perihelion precession per century in Eq. \eqref{e_precession_final}, where we evaluate each term numerically for Mercury, the planet with the most accurately measured perihelion shift and the strongest relativistic effects in the solar system. We adopt the following physical parameters: Mercury's semi-major axis $ a = 5.791 \times 10^{10} \, \text{m} $, and orbital eccentricity $ e = 0.205630 $. We also use the loop coefficients that are taken from the graviton-dominated values (see Eq. \eqref{graviton}). Furthermore, in one century, Mercury's number of revolutions is $415.2$.

For the first term of Eq. \eqref{e_precession_final}, which is the classical general relativistic contribution, yields the well-known result $ \Delta \phi_{\text{GR}} \approx 43.00$ arcsecond/century, consistent with both theory and observational data. The first quantum correction term gives $2.467 \times 10^{-9}$ arcsecond/century, while that of the second correction gives $-7.155 \times 10^{-10}$ arcsecond/century. Taken together, the net quantum correction to the perihelion shift of Mercury is approximately
\begin{equation}
    \Delta \phi_{\text{QG}} \approx 1.752 \times 10^{-9}\text{ arcsecond}/\text{century},
\end{equation}
which is over six orders of magnitude below the precision of current ephemeris-based measurements, whose uncertainties remain at the level of $ \sim 10^{-3} $ arcsecond per century or larger.

Although such a minute correction lies far beyond current observational capabilities, this analysis demonstrates that the quantum-corrected EFT formalism produces theoretically well-defined and analytically distinct predictions. The separation of terms involving $ \hat{\alpha} $ and $ \hat{\beta} $ offers a framework for future constraints, should the precision of planetary tracking or interplanetary ranging experiments improve by several orders of magnitude. More importantly, it illustrates that even in weak-field limits, effective quantum gravity introduces qualitatively new structures into gravitational observables, breaking the classical equivalence between matter-supported and vacuum spacetimes.

\subsection{Shapiro-Time delay}
The Shapiro time delay, also known as gravitational time delay, is the increase in the round-trip travel time of a light signal as it passes through the gravitational field of a massive body \cite{Shapiro:1964uw}. First predicted by Irwin Shapiro in 1964, it has since become one of the cornerstones of weak-field tests of GR, most notably confirmed through radar echo measurements between Earth and other planets such as Venus or Mercury.

We derive here the Shapiro delay for a radar signal propagating along a nearly radial path from an observer at coordinate $ r_1 $, grazing a central mass at minimal radial distance $ r_0 $, and being reflected back from $ r_2 $. The trajectory lies in the equatorial plane $ \theta = \pi/2 $, and the signal follows a null geodesic $ ds^2 = 0 $.

From the line element in Eq. \eqref{metric2}, the null condition yields:
\begin{equation}
0 = -A(r) \, dt^2 + B(r) \, dr^2 + C(r) \, d\phi^2,
\end{equation}
for motion in the equatorial plane. Since the light ray is radial at closest approach and travels with negligible angular change near $ r_1, r_2 $, the deflection is small and we may set $ d\phi \approx 0 $. Then the null condition simplifies to:
\begin{equation}
dt = \sqrt{\frac{B(r)}{A(r)}}dz,
\end{equation}
where, for a nearly straight-line trajectory, it’s better to rewrite in terms of \(dz\), where \(z\) is along the straight line and:
\begin{equation}
    r(z) = \sqrt{z^2 + r_0^2}, \quad dr = \frac{z}{\sqrt{z^2 + r_0^2}} dz.
\end{equation}
Now, we compute
\begin{equation}
    \Delta t = \int_{-z_1}^{z_2} \left[ 1 + \frac{2GM}{r(z)} + \frac{G^2 M}{2 r(z)^3} (\hat{\alpha} + \hat{\beta}) \right] dz,
\end{equation}
where we applied Taylor expansion. Subtracting the flat space time \(t_0 = z_1 + z_2\), we get the time delay:
\begin{equation}
    \Delta t_{\text{Shapiro}} = 2GM \int_{-z_1}^{z_2} \frac{dz}{r(z)} + \frac{G^2 M}{2} (\hat{\alpha} + \hat{\beta}) \int_{-z_1}^{z_2} \frac{dz}{r(z)^3}.
\end{equation}
With $r(z) = \sqrt{z^2 + r_0^2}$ and doing large-$r$ approximation, we arrive at
\begin{equation}
    \Delta t_{\text{Shapiro}} = 4GM \ln \left( \frac{4 r_1 r_2}{r_0^2} \right) + \frac{G^2 M}{r_0^2} (\hat{\alpha} + \hat{\beta}).
\end{equation}
Here, the logarithmic term is the classical Shapiro delay, while the second term is the quantum correction, which decays as $ 1/r_0^2 $. This correction becomes appreciable only for small impact parameters and is suppressed by an additional factor of $ G $, as expected from a loop-level effect.

To assess the physical relevance of the quantum correction to the Shapiro time delay, we consider a typical solar system radar experiment, such as a radar echo transmitted from Earth, reflected off a spacecraft near another planet (e.g., Mercury or Mars), and received back on Earth. The path of the signal passes near the Sun, resulting in a delay relative to the expected signal time in flat spacetime. The delay depends strongly on the closest approach distance, known as the impact parameter, typically approximated as the solar radius $ r_0 \approx R_\odot $.

We now compute both contributions numerically for the case of (a) radar sent from Earth ($ r_1 \approx 1 \, \text{AU} $), (b) reflected from a probe at Mars ($ r_2 \approx 1.5 \, \text{AU} $), and (c) passing near the Sun ($ r_0 \approx R_\odot $). Note that $1 \text{AU} = 1.495978707 \times 10^{11} \, \text{m}$. For the classical Shapiro delay, we find $\Delta t_{\text{GR}} \sim 247 \mu\text{s}$, which is consistent with the typical Shapiro delay observed in radar-ranging experiments (on the order of several tens of microseconds to a few seconds, depending on the geometry). Now, for the quantum correction term, we find
\begin{equation}
\Delta t_{\text{QG}} = 2.719\times 10^{-23} \, \text{s},
\end{equation}
where we have used the previously estimated graviton loop values for $\hat{\alpha}$ and $\hat{\beta}$. The quantum correction is smaller than the classical Shapiro delay by approximately:
\begin{equation}
\frac{|\Delta t_{\text{QG}}|}{\Delta t_{\text{GR}}} \approx \frac{2.719 \times 10^{-23}}{2.47 \times 10^{-4}} \approx 1.10 \times 10^{-19}.
\end{equation}
Such a level of correction is well beyond the reach of current radar and timing experiments. For references, the Cassini tracking experiments achieved timing accuracy of order $ 10^{-12} \, \text{s} $. Future missions with atomic clocks in deep space may improve this sensitivity. But the quantum correction remains way smaller than even the best experimental resolution available today.

Nevertheless, the calculation demonstrates that the Shapiro delay is sensitive to the combination $ \hat{\alpha} + \hat{\beta} $, distinct from the redshift (which probes $ \hat{\alpha} $ alone). Hence, a comprehensive testing program involving multiple classical observables could, in principle, over-constrain the quantum EFT parameters if sufficient experimental sensitivity were attained.

\subsection{Gravitational Redshift}
The gravitational redshift, one of the classical experimental tests of GR, is the frequency shift experienced by electromagnetic radiation when propagating from a region of stronger gravitational potential to a region of weaker potential \cite{Einstein:1911vc,Nicolini:2009gw,Cardenas:2021eri}. In the weak-field approximation, this redshift is governed by the temporal component of the spacetime metric. In this section, we derive the gravitational redshift formula, including quantum gravitational corrections to the exterior metric of a static, spherically symmetric mass distribution, as obtained from the EFT of gravity.

We consider a photon emitted at a radial coordinate $ r_e $, and received at $ r_r $, with both emitter and receiver following static worldlines in the given spacetime. The observed frequency shift is determined by the ratio of proper time intervals at the respective locations. For a static observer at a fixed spatial coordinate, the proper time interval $ d\tau $ is related to the coordinate time $ dt $ by:
\begin{equation}
d\tau = \sqrt{A(r)} \, dt.
\end{equation}
Let $ \nu_e $ and $ \nu_r $ denote the proper frequencies of successive wave crests of a photon measured by observers at $ r_e $ and $ r_r $, respectively. Since the coordinate time interval $ dt $ between wave crests is the same for both observers, it follows that
\begin{equation}
\frac{\nu_r}{\nu_e} = \frac{\sqrt{A(r_r)}}{\sqrt{A(r_e)}}.
\end{equation}
The gravitational redshift $ z $, defined as
\begin{equation}
z \equiv \frac{\nu_e - \nu_r}{\nu_r} = \left[ \frac{A(r_r)}{A(r_e)} \right]^{1/2} - 1,
\end{equation}
quantifies the fractional frequency change experienced by the photon.

In many experimental scenarios, the receiver is located far from the gravitational source, such that $ r_r \to \infty $. In this limit, the spacetime becomes asymptotically flat and $ A(r_r) \to 1 $, so the redshift reduces to
\begin{equation}
z = \left[A(r_e)\right]^{-1/2} - 1.
\end{equation}
Using the quantum-corrected expression for the temporal component of the metric in Eq. \eqref{metric1},
\begin{equation}
A(r_e) = 1 - \frac{2GM}{r_e} - \frac{\hat{\alpha} G^2 M}{r_e^3},
\end{equation}
we expand the inverse square root to first order in $ GM/r_e $ and $ G^2 M/r_e^3 $. Let
\begin{eqnarray}
\epsilon \equiv -\left( \frac{2GM}{r_e} + \frac{\hat{\alpha} G^2 M}{r_e^3} \right),
\end{eqnarray}
so that $ A(r_e) = 1 + \epsilon $. Then,
\begin{eqnarray}
\left[A(r_e)\right]^{-1/2} = (1 + \epsilon)^{-1/2} 
\approx 1 - \frac{1}{2} \epsilon + \frac{3}{8} \epsilon^2 + \cdots\approx 1 + \frac{GM}{r_e} + \frac{\hat{\alpha} G^2 M}{2 r_e^3} + \mathcal{O}(G^3).
\end{eqnarray}
Consequently, the redshift becomes:
\begin{equation}
z = \frac{GM}{r_e} + \frac{\hat{\alpha} G^2 M}{2 r_e^3} + \mathcal{O}(G^3).
\end{equation}

The leading-order term $ z_{\text{GR}} = GM/r_e $ recovers the well-known prediction of GR for the gravitational redshift near a spherical body. The second term represents the leading quantum correction, proportional to $ \hat{\alpha} $, which enters at order $ \mathcal{O}(G^2) $. The correction decays rapidly with distance as $ 1/r^3 $, consistent with its origin in higher-order curvature corrections to the gravitational potential.

For an emitter located near the surface of the Sun, where we used typical values of the parameters concerned, we obtain the classical GR prediction as $(G=1)$
\begin{equation}
    z_{\text{GR}} = \frac{M_\odot}{R_\odot} \approx 2.121 \times 10^{-6},
\end{equation}
which is consistent with empirical measurements of the solar redshift, e.g., via the Pound-Rebka experiment or spectroscopic solar lines. Now, consider the quantum gravitational correction
\begin{equation}
    z_{\text{QG}} = \frac{\hat{\alpha} G^2 M_{\odot}}{2 R_{\odot}^3} \approx 9.886 \times 10^{-24},
\end{equation}
The quantum gravitational correction is thus eighteen orders of magnitude smaller than the classical GR redshift. This renders the correction unobservable with current precision spectroscopy, where experimental uncertainties are typically no better than $ \sim 10^{-8} $ in solar redshift measurements.

Nevertheless, this term is not without significance. The fact that the quantum correction appears explicitly at order $ \mathcal{O}(G^2) $, demonstrates that gravitational redshift may, in principle, serve as a theoretical probe of the quantum structure of spacetime. If future advances in atomic clock precision or space-based frequency standards were to push sensitivities into the $ 10^{-24} $ regime or beyond, this correction would enter the threshold of empirical testability.

Moreover, the suppression of the quantum correction as $ \sim 1/r^3 $ implies that these effects become more prominent near compact objects. While inaccessible in solar-system contexts, such corrections may be relevant in theoretical studies of Planck-scale compact objects or near-horizon quantum gravity phenomenology.

\section{Propagation of scalar waves and QNMs}\label{perturbations}
%%%%%%%%%%%%%%%%%%%%%%%%%%%%%%%%%%%%%%%

Investigating wave propagation within a given spacetime geometry is an effective approach for probing fundamental features of that geometry by analyzing the scattering and absorption behavior of the waves. Such studies provide critical insight into the physical viability and stability of these spacetime backgrounds, particularly under linear perturbations. Stability against these perturbations is a fundamental requirement for any semiclassical spacetime to be considered a realistic description of astrophysical phenomena, such as black holes or compact stellar objects \cite{Nollert:1999ji,Ferrari:2007dd,Berti:2009kk,Konoplya:2011qq}.

%These frequencies depend only on the geometry around the light-ring of the classical black hole, and describe  the early ringdown stage in gravitational wave observations of binary mergers. While the late ringdown stage is expected to be described by the proper QNM frequencies, the calculation of these \black{is} out of the scope of the present paper.

Let us now explicitly examine the evolution of a massless scalar field propagating on a general static and spherically symmetric spacetime described by the metric: $ds^2=g_{tt}dt^2 + g_{rr}dr^2+ r^2 d\Omega^2$. 
The dynamics of this scalar field are governed by the Klein-Gordon equation,
\be
\Box \phi = 0\,,
\ee
where the D'Alembertian operator in curved spacetime takes the form:
\be
\Box \phi = \frac{1}{\sqrt{-g}}\partial_\mu\left(\sqrt{-g}\,g^{\mu\nu}\partial_\nu \phi\right)\,.
\ee

Given the static and spherically symmetric symmetry of the metric, we seek solutions of the scalar field equation in a separable form:
\be
\phi_{\omega l m}(t,r,\theta,\psi) = \frac{1}{r}\,e^{-i \omega t}\,Y_{lm}(\theta,\psi)\,\Psi_{\omega l}(r)\,,
\ee
where $Y_{lm}$ are spherical harmonics encoding angular dependence. Substituting this ansatz into the Klein-Gordon equation and performing a careful decoupling of variables, we derive an ordinary differential equation for the radial function $\Psi_{\omega l}(r)$:
\be
F^2(r)\,\frac{d^2\Psi_{\omega l}}{dr^2}+F(r)F'(r)\,\frac{d\Psi_{\omega l}}{dr}+\left(\omega^2-V_l(r)\right)\Psi_{\omega l}=0\,,
\ee
where the radial-dependent function $F(r)$ and the effective potential $V_l(r)$ are explicitly defined as \cite{Beltran-Palau:2022nec}:
\be
F(r) = \sqrt{-\frac{g_{tt}(r)}{g_{rr}(r)}}\,,
\ee
\be
V_l(r) = -g_{tt}(r)\frac{l(l+1)}{r^2}+\frac{F(r)\,F'(r)}{r}\,.
\ee

To simplify the structure of this equation, we introduce a generalized tortoise coordinate $r^*$ defined via:
\be
\frac{d}{dr^*}=F(r)\frac{d}{dr}\,.
\ee
Under this transformation, the radial equation assumes a form identical to the well-known Regge-Wheeler equation:
\be
\frac{d^2\Psi_{\omega l}}{dr^{*2}} + \left(\omega^2 - V_l(r^*)\right)\Psi_{\omega l}=0\,.
\ee

As a consistency check, applying this procedure to the Schwarzschild spacetime precisely reproduces the familiar standard results, highlighting the robustness and generality of our approach.

The effective potential $V_l(r^*)$ may thus be explicitly expressed as follows:
\bea
V(r)=\frac{l (l+1) \left(-\alpha  G^2 M-2 G M r^2+r^3\right)}{r^5}+\frac{(2 G M-r) (4 G M-3 r)}{2 \left(\beta  G^2 M (r-2 G M)+r^4\right)}-\frac{r^2 (3 r-8 G M) \left(-\alpha  G^2 M-2 G M r^2+r^3\right)}{2 \left(\beta  G^2 M (r-2 G M)+r^4\right)^2}.
\eea

\begin{figure}[htp!]
\centering
\includegraphics[scale=0.6]{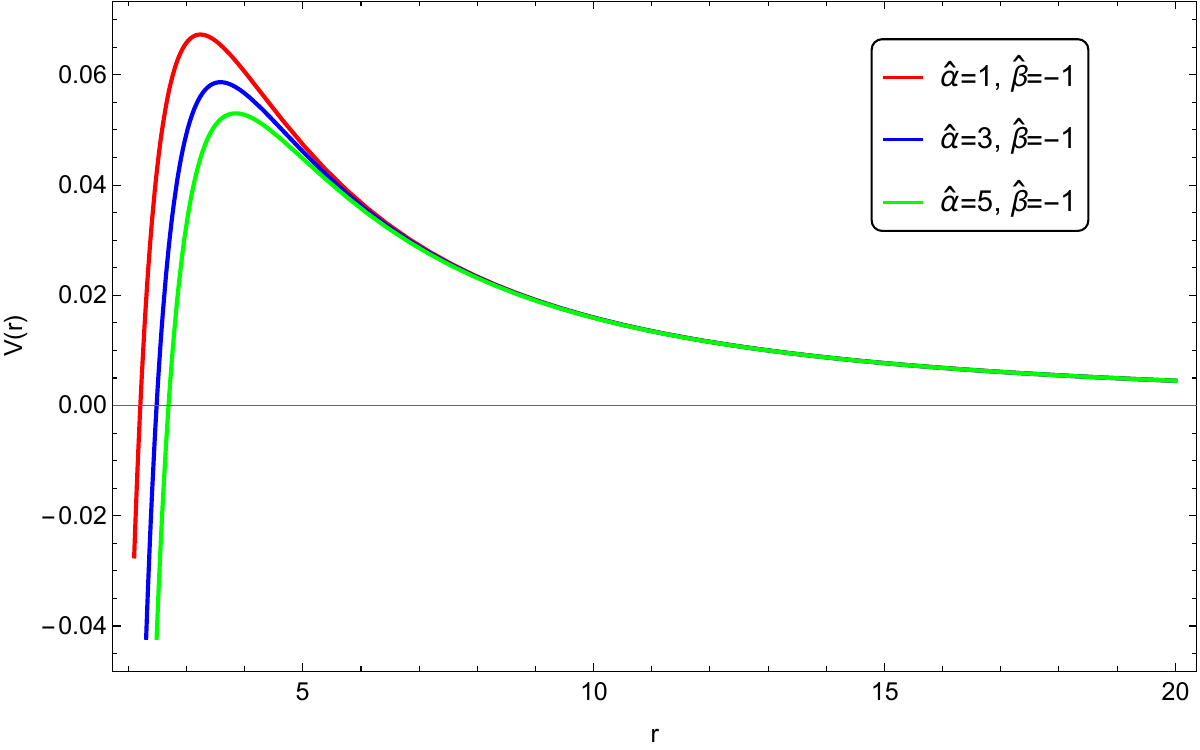} 
\caption{
Effective potential barrier for massless scalar perturbations against the radial coordinate for the parameters shown in the panels. To highlight the effect of parameter $\hat{\alpha}$, we choose large values for it.
}
\label{fig:0} 	
\end{figure}

In Fig. \ref{fig:0}, we present the effective potential barrier \( V(r) \) for massless scalar field perturbations as a function of the radial coordinate \( r \), considering distinct values of the model parameters \( \hat{\alpha} \) and \( \hat{\beta} \). Notably, as \( \hat{\alpha} \) increases from 1 to 5 with fixed \( \hat{\beta}=-1 \), the height and width of the potential barrier visibly increase, indicating a stronger potential well that significantly impacts the propagation of scalar waves around the black hole. Such an increment suggests enhanced stability against scalar perturbations for larger \( \hat{\alpha} \) values. Moreover, the shape and magnitude of \( V(r) \) imply that scalar fields encounter a higher potential barrier, thus affecting the quasinormal mode spectrum and scattering properties.

The quasinormal-mode problem for a wide class of black holes reduces to a one-dimensional Schrodinger-like equation,  
\begin{eqnarray}
\frac{d^2\Psi}{dx^2} + \bigl(\omega^2 - V(x)\bigr)\,\Psi = 0,
\end{eqnarray}
where \(x\) is the tortoise coordinate and \(V(x)\) is a single-peak potential vanishing at the horizon and spatial infinity.  In the WKB approach, one matches the approximate WKB solutions in the asymptotic regions to a Taylor expansion of the potential about its maximum, obtaining at the \(k\)-th order the quantization condition \cite{Konoplya:2003ii,Konoplya:2011qq,Konoplya:2019hlu}  
\begin{eqnarray}
V_0 + \sum_{n=2}^{k}A_n(K^2)\;-\;
i\,K\,\sqrt{-2V_2}\,\Bigl(1+\sum_{\substack{n=3\\n\ \mathrm{odd}}}^{k}A_n(K^2)\Bigr)
=0,
\end{eqnarray}
with \(K=n+1/2\) the overtone index and \(V_j\) the \(j\)th derivative of \(V\) at its peak.  This yields an efficient and useful procedure for scanning quasinormal frequencies and greybody factors.

\begin{table}[ht]
\caption{WKB 6 orders QNMs: real and imaginary parts of $\omega$ for $l=1$.}
\centering
\begin{tabular}{c c c}
\toprule
Order & $\Re(\omega)$      & $\Im(\omega)$       \\
\midrule
  6 & 0.2771334716186022 & -0.0966991544955449 \\
  5 & 0.2767036135378085 & -0.0968493762885831 \\
 4 & 0.2766953725064447 & -0.0968258287291509 \\
 3 & 0.2745158737858119 & -0.0975945703211363 \\
 2 & 0.2789264388103875 & -0.1093878122243252 \\
 1 & 0.3176304698850302 & -0.0960586461495153 \\
\bottomrule
\end{tabular}
%\caption{WKB 6 orders QNMs: real and imaginary parts of $\omega$ for $l=1$.}
\label{tab:QNM6}
\end{table}

\begin{figure}[htp!]
\centering
\includegraphics[scale=0.6]{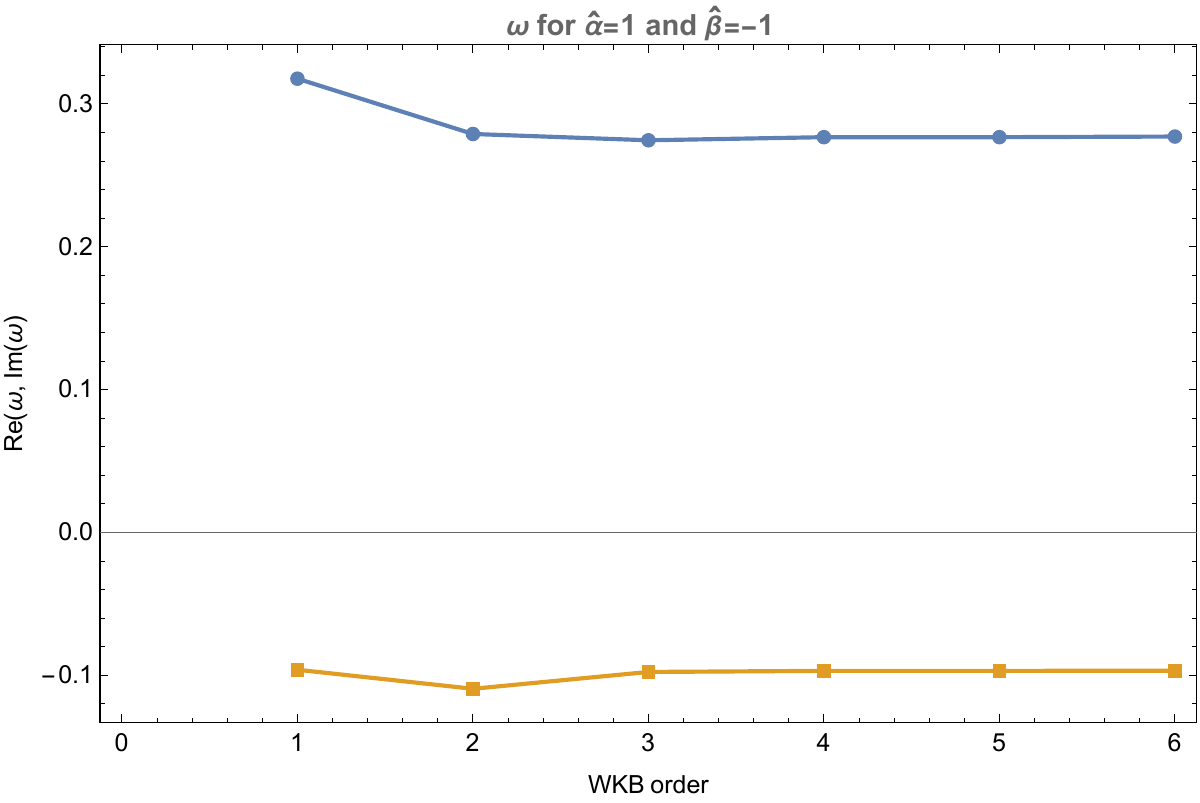} 
\caption{
The behaviors of real and imaginary parts of the QNMs with different WKB orders under
massless scalar perturbations for the parameters shown in the panel.}
\label{fig:1} 	
\end{figure}

\begin{table}[ht]
\caption{Quasinormal mode frequencies \(\omega\) from WKB 13 order: real and imaginary parts for $l=1$.}
\centering
\begin{tabular}{c c c}
\toprule
Order & $\Re(\omega)$      & $\Im(\omega)$       \\
\midrule
13 & 0.2707006682618470 &  0.2703096026896804 \\
12 & 0.1971526467396140 & -0.1966153494142238 \\
11 & 0.3040919811772292 & -0.1274720772859067 \\
10 & 0.2893386252877738 & -0.0865681093400488 \\
9 & 0.2724071761478131 & -0.0919487441719237 \\
8 & 0.2747815093493479 & -0.0987612262898448 \\
  7 & 0.2775153819438508 & -0.0977883050482830 \\
6 & 0.2771334716186022 & -0.0966991544955449 \\
5 & 0.2767036135378085 & -0.0968493762885831 \\
4 & 0.2766953725064447 & -0.0968258287291509 \\
3 & 0.2745158737858119 & -0.0975945703211363 \\
2 & 0.2789264388103875 & -0.1093878122243252 \\
1 & 0.3176304698850302 & -0.0960586461495153 \\
\bottomrule
\end{tabular}
%\caption{Quasinormal mode frequencies \(\omega\) from WKB 13 order: real and %imaginary parts for $l=1$.}
\label{tab:QNM13}
\end{table}

\begin{figure}[htp!]
\centering
\includegraphics[scale=0.6]{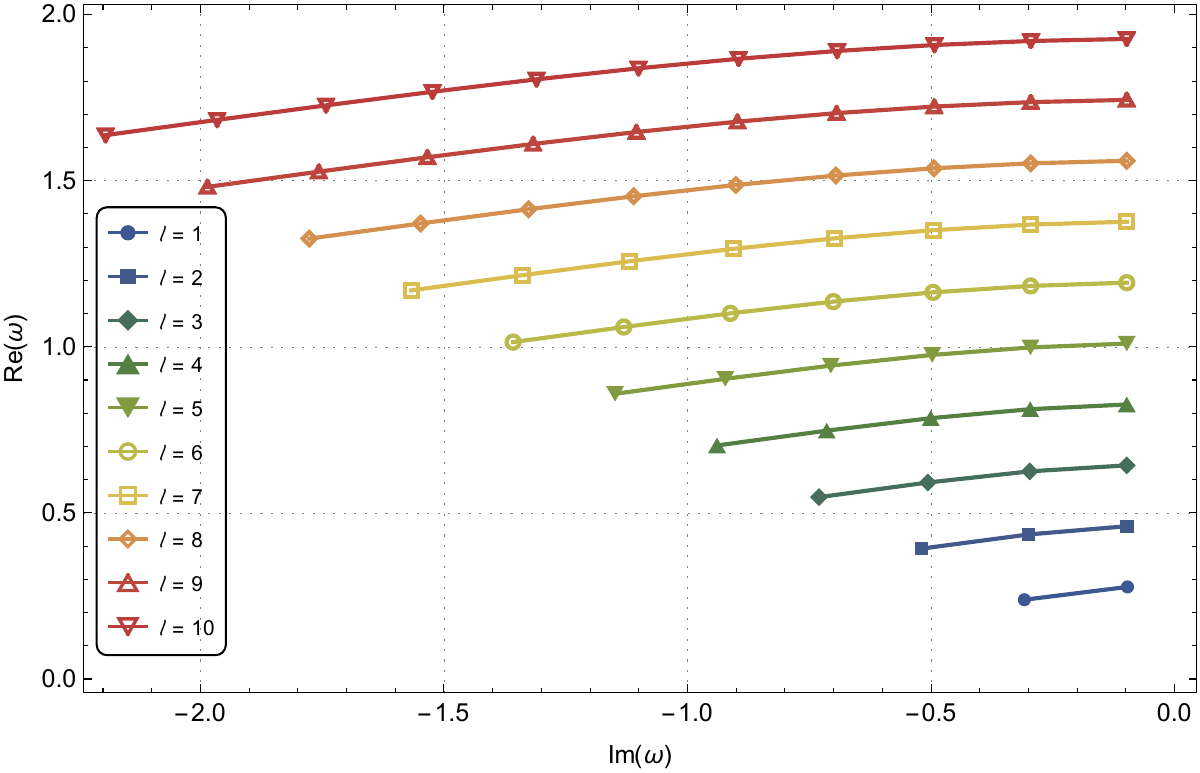} 
\caption{
The QNM spectrum for $\hat{\alpha}=M=G=1$ and $\hat{\beta}=-1$. We show the cases in which $l$ is \{1-10\}.
}
\label{fig:2} 	
\end{figure}

\begin{table}[htbp]
\centering
\footnotesize
\caption{Fundamental QNM $\omega$ at different WKB orders, with estimated error$^{a}$.}
\label{tab:QNMerror}
\begin{tabular}{c l l}
\toprule
Order & $\omega$ & Error \\
\midrule
12 & $0.197153 - 0.196615\,i$ & $0.1996$ \\
11 & $0.304092 - 0.127472\,i$ & $0.07178$ \\
10 & $0.289339 - 0.086568\,i$ & $0.02380$ \\
 9 & $0.272407 - 0.091949\,i$ & $0.009494$ \\
 8 & $0.274782 - 0.098761\,i$ & $0.003879$ \\
 7 & $0.277515 - 0.097788\,i$ & $0.001564$ \\
 6 & $0.277133 - 0.096699\,i$ & $0.000621$ \\
 5 & $0.276704 - 0.096849\,i$ & $0.000228$ \\
 4 & $0.276695 - 0.096826\,i$ & $0.001156$ \\
 3 & $0.274516 - 0.097595\,i$ & $0.006379$ \\
\bottomrule
\end{tabular}

\vspace{2pt}
\footnotesize
\raggedright
$^{a}$Error is estimated by comparing WKB formula values of successive orders.
\end{table}

\begin{table}[htbp]
\centering
\footnotesize
\caption{Fundamental QNM $\omega$ using Pade approximation and estimated error at each WKB order.}
\label{tab:QNMpade}
\begin{tabular}{c l l}
\toprule
$n$ & $\omega$ & Error \\
\midrule
 1  & $0.291014 - 0.088009\,i$ & $2.78\times10^{-2}$ \\
 2  & $0.278057 - 0.094401\,i$ & $2.63\times10^{-3}$ \\
 3  & $0.276398 - 0.096256\,i$ & $4.68\times10^{-7}$ \\
 4  & $0.276781 - 0.097031\,i$ & $2.16\times10^{-4}$ \\
 5  & $0.276789 - 0.096862\,i$ & $8.68\times10^{-5}$ \\
 6  & $0.276923 - 0.096969\,i$ & $1.21\times10^{-4}$ \\
 7  & $0.276804 - 0.096958\,i$ & $1.99\times10^{-5}$ \\
 8  & $0.276809 - 0.096969\,i$ & $1.60\times10^{-5}$ \\
 9  & $0.276802 - 0.096954\,i$ & $6.23\times10^{-6}$ \\
10  & $0.276804 - 0.096966\,i$ & $1.92\times10^{-5}$ \\
11  & $0.276802 - 0.096944\,i$ & $1.29\times10^{-6}$ \\
12  & $0.276803 - 0.096943\,i$ & $3.24\times10^{-7}$ \\
13  & $0.276800 - 0.096942\,i$ & $4.46\times10^{-6}$ \\
\bottomrule
\end{tabular}
\end{table}

\begin{table}[htbp]
\centering
\footnotesize
\caption{WKB quasinormal-mode frequency $\omega$ as a function of the parameter $\hat\alpha$.}
\label{tab:alpha_vs_wkb}
\begin{tabular}{r l}
\toprule
$\hat\alpha$ & $\omega_{\rm WKB}$ \\
\midrule
  1  & $0.277133 - 0.096699\,i$ \\
  3  & $0.254790 - 0.092804\,i$ \\
  5  & $0.239316 - 0.088819\,i$ \\
  7  & $0.227462 - 0.085209\,i$ \\
  9  & $0.217857 - 0.081983\,i$ \\
 11  & $0.209789 - 0.079090\,i$ \\
 13  & $0.202840 - 0.076477\,i$ \\
 15  & $0.196741 - 0.074100\,i$ \\
 17  & $0.191309 - 0.071923\,i$ \\
 19  & $0.186417 - 0.069917\,i$ \\
 21  & $0.181968 - 0.068058\,i$ \\
\bottomrule
\end{tabular}
\end{table}

\begin{figure}[htp!]
\centering
\includegraphics[scale=0.6]{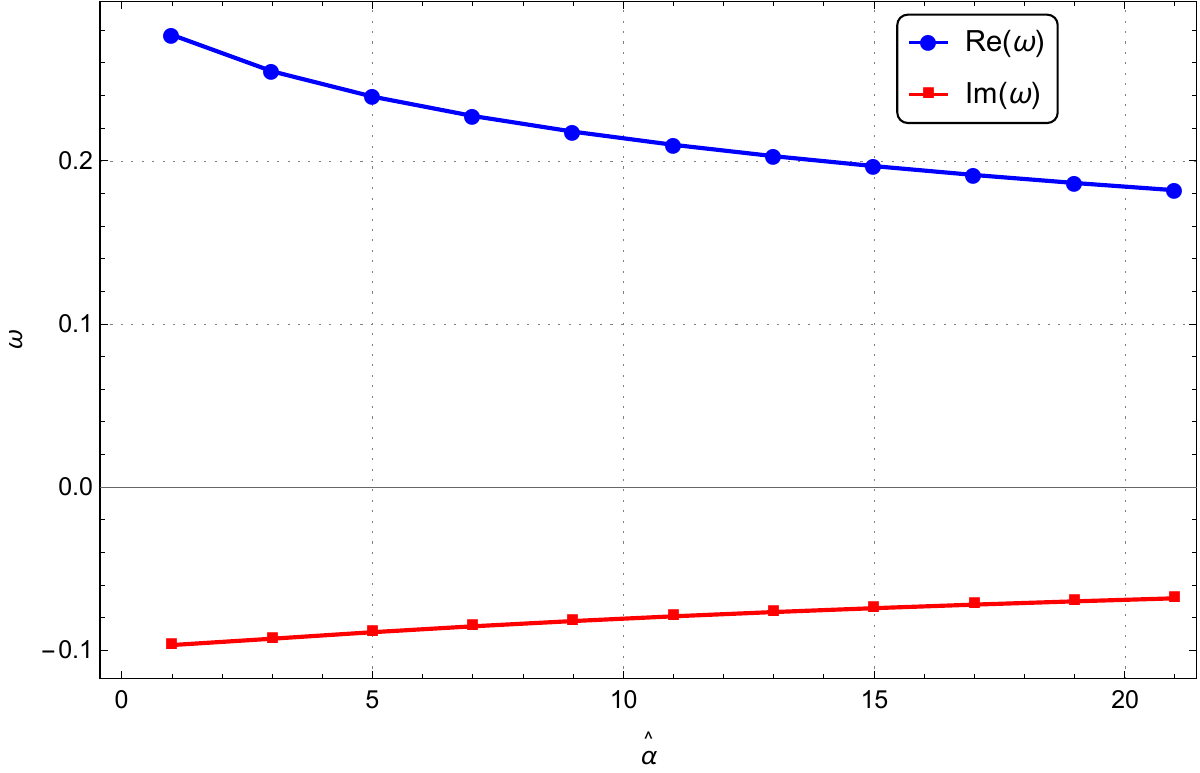} 
\caption{
The QNM frequencies for different values of $\hat{\alpha}$ parameters.
}
\label{fig:3} 	
\end{figure}

\begin{figure}[htp!]
\centering
\includegraphics[scale=0.6]{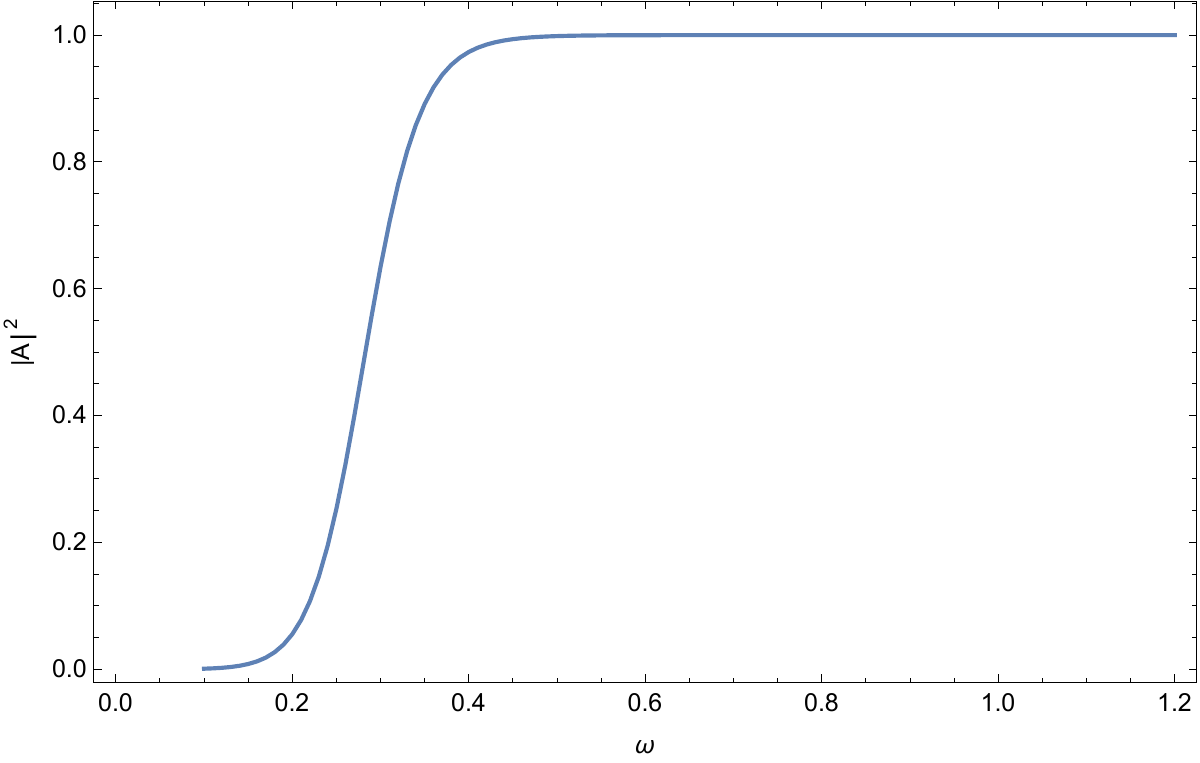} 
\caption{
The graybody factor for $\hat{\alpha}=M=G=1$ and $\hat{\beta}=-1$.
}
\label{fig:4} 	
\end{figure}

\begin{figure}[htp!]
\centering
\includegraphics[scale=0.6]{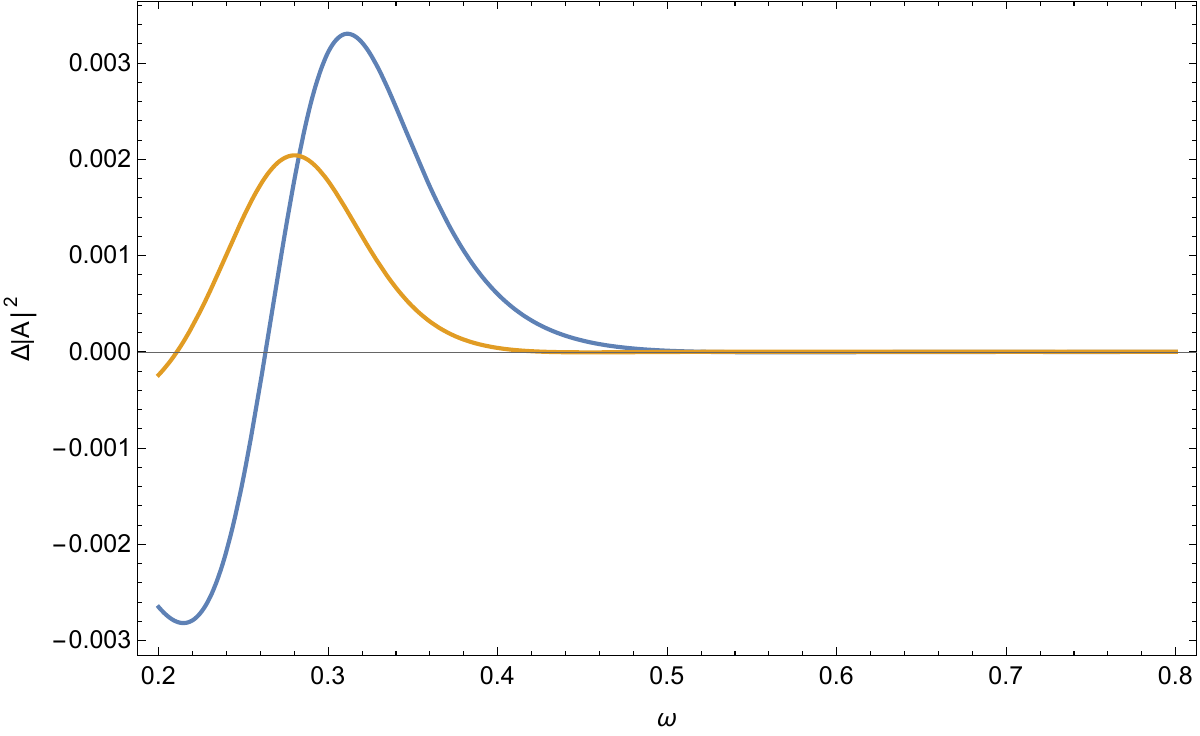} 
\caption{
The graybody factor different between different orders $\hat{\alpha}=M=G=1$ and $\hat{\beta}=-1$. We show the cases in which $l$ is \{1-10\}.
}
\label{fig:5} 	
\end{figure}

The provided figures \ref{fig:1}-\ref{fig:3} and tables \ref{tab:QNM6}-\ref{tab:QNMpade}  illustrate important characteristics of the QNMs (QNMs) and the corresponding greybody factors. Specifically, plots of Fig.\ref{fig:3} and Table \ref{tab:alpha_vs_wkb} the real and imaginary parts of the frequency (\(\omega\)) as functions of the parameter \(\hat{\alpha}\) depict clear trends indicating how varying the coupling affects the stability and oscillation frequency of perturbations. The real part of the frequency (Re(\(\omega\))) shows a monotonically increasing behavior, suggesting that the oscillation frequencies of perturbations rise as \(\hat{\alpha}\) increases. Conversely, the imaginary part (Im(\(\omega\))) displays a decreasing trend, becoming more negative with increasing \(\hat{\alpha}\). This indicates enhanced damping, implying stronger stability and faster decay of perturbations at higher \(\hat{\alpha}\).

In the greybody factor plots Fig. \ref{fig:4}-\ref{fig:5}, we observe how effectively the black hole emits radiation at various frequencies. The depicted transmission amplitude (\(|A|^2\)) highlights frequency-dependent peaks, signifying optimal emission frequencies for particle modes. The shape and magnitude of these peaks provide critical information regarding the transparency of the black hole spacetime to scalar perturbations. Overall, the presented data underscores a rich interplay between coupling parameters and the QNMs, offering insights into stability analysis and observational predictions for gravitational wave signals in modified gravity scenarios.

\section{Conclusion} \label{conc}
The present study reinforces the compelling assertion that quantum gravitational effects imprint discernible deviations on classical spacetime geometries, particularly in scenarios involving matter-supported configurations like stars. By employing the Gauss-Bonnet theorem and analyzing geodesic deflection in the weak field limit, we have provided explicit analytical expressions that exhibit how the quantum correction parameters $ \hat{\alpha} $ and $ \hat{\beta} $ influence both null and time-like trajectories. These corrections, while exceedingly small under solar system conditions, introduce unique fingerprints that differentiate quantum-corrected stellar spacetimes from classical black hole geometries.

An essential implication of this analysis, aligning with the foundational work in Ref. \cite{Calmet:2019eof}, is the nontrivial realization that vacuum solutions — particularly the Schwarzschild exterior — remain unaltered under these quantum corrections at second order in curvature. In contrast, the stellar spacetimes, under their nonzero stress-energy content, inherit non-local, curvature-squared modifications that do not vanish asymptotically but are confined to subleading terms. This stark dichotomy calls into question the adequacy of using classical black hole metrics to model astrophysical systems that emerge from collapse scenarios and which, in principle, retain quantum information imprints from their progenitor matter configuration.

Furthermore, the results on light deflection and perihelion precession extend the narrative beyond mere conceptual insight, offering a concrete pathway for constraining EFT parameters using high-precision observational probes. Although current instrumentation lacks the sensitivity to detect the predicted $ \sim 10^{-9} $ arcsecond/century quantum shifts, future technological advances, particularly in astrometric and time-delay missions, may eventually access this regime.

Taken together, this work supports the broader claim that quantum-corrected metrics, as derived via effective actions respecting general covariance, are not merely theoretical artifacts but offer a more refined lens for modeling compact objects in regimes where classical and quantum gravity overlap. In particular, the operational distinction between stars and black holes, previously blurred in classical GR, is reinstated under quantum considerations, potentially bearing on foundational issues such as black hole information recovery, the nature of horizon formation, and the definition of entropy in non-vacuum geometries.

A final remark regarding the black hole shadow, which we have not analyzed in this paper. The classical notion of a black hole shadow, rooted in the existence of a photon sphere and the geometric tracing of null geodesics in a smooth spacetime, loses operational and conceptual meaning in the Planck-scale regime. At such scales, the semiclassical framework that defines light propagation, and by extension, shadow formation, ceases to be valid. The geometric optics limit, which underpins the visibility of shadows in GR, is fundamentally compromised by quantum fluctuations in both spacetime and the propagation of fields. As a result, a Planck-mass black hole does not admit a well-defined or observationally accessible shadow, and any residual light-trapping structure, if it exists at all, would be subject to extreme quantum ambiguity.

Nonetheless, the idea of a shadow may persist in a generalized, probabilistic, or expectation-valued sense within quantum gravity frameworks. Theoretical constructions involving non-singular black holes, generalized uncertainty principles, or effective field-theoretic modifications to the metric may yield partial analogues of photon spheres. These possibilities invite a rethinking of what constitutes a “shadow” in quantum spacetime, pointing toward a more fundamental and statistical description of light propagation in the deep quantum gravitational regime. Such considerations represent a fertile ground for extending classical black hole phenomenology into the Planck domain.

Last, we have thoroughly analyzed the propagation of massless scalar waves and the corresponding quasinormal mode (QNM) spectrum in a general static and spherically symmetric spacetime. Employing the WKB approximation up to high orders, supplemented by Pade resummation techniques, our study provides robust quantification of the scalar perturbation stability and emission properties. Our numerical results demonstrate that as the coupling parameter increases, the effective potential barrier strengthens, elevating both the oscillation frequencies and the damping rates of perturbations, thus reinforcing the spacetime's stability. Additionally, the greybody factor analysis elucidates the frequency-dependent transparency of the spacetime, highlighting the impact of the coupling parameters on scalar wave emissions. These findings offer valuable theoretical insights relevant to future gravitational wave detections and astrophysical observations within modified gravity contexts.

\acknowledgements

R. P., A. \"O. and G. L. would like to acknowledge networking support of the COST Action CA21106 - COSMIC WISPers in the Dark Universe: Theory, astrophysics and experiments (CosmicWISPers), the COST Action CA22113 - Fundamental challenges in theoretical physics (THEORY-CHALLENGES), the COST Action CA21136 - Addressing observational tensions in cosmology with systematics and fundamental physics (CosmoVerse), the COST Action CA23130 - Bridging high and low energies in search of quantum gravity (BridgeQG), and the COST Action CA23115 - Relativistic Quantum Information (RQI) funded by COST (European Cooperation in Science and Technology). A. \"O. also thanks to EMU, TUBITAK, ULAKBIM (Turkiye) and SCOAP3 (Switzerland) for their support.

\bibliography{ref}

\end{document}